\documentclass[11pt]{scrartcl} 
\usepackage[boxed,ruled,vlined,linesnumbered,noresetcount]{algorithm2e}
\usepackage{amsmath,latexsym,amsopn,amsfonts,amssymb} 
\usepackage{soul}
\usepackage{enumitem}
\usepackage{graphicx,xspace}

\newtheorem{remark}{Remark}[section]
\newtheorem{theorem}{Theorem}[section]
\newtheorem{lemma}[theorem]{Lemma}
\newtheorem{definition}{Definition}[section]

\newtheorem{claim}[theorem]{Claim}

\newcommand\bull{{\operatorname{-\xspace}}}
\newcommand{\blitza}{\text{\usefont{U}{ulsy}{m}{n}\symbol{'011}}}
\usepackage{color}
\newcommand{\EMS}[1]{[\textcolor{black}{EMS: #1}]} 

\newcommand{\ems}[1]{\textcolor{black}{#1}}
\newcommand{\emsA}[1]{\textcolor{black}{#1}}
\newcommand{\rd}[1]{\textcolor{black}{#1}}
\newcommand{\algSize}{normalsize} 
\newcommand{\bigO}{\mathcal{O}\xspace}
\newcommand{\remove}[1]{}
\newcommand{\reduce}[1]{} 

\newcommand{\Correct}{\mathit{Correct}\xspace} 
\newcommand{\init}{\texttt{init}\xspace} 
\newcommand{\valid}{\texttt{valid}\xspace} 
\newcommand{\initE}{\emph{\texttt{init}}\xspace} 
\newcommand{\validE}{\emph{\texttt{valid}}\xspace} 
\newcommand{\echo}{\texttt{echo}\xspace} 
\newcommand{\ready}{\texttt{ready}\xspace} 
\newcommand{\etal}{\emph{et al.}\xspace}
\newcommand{\eg}{\emph{e.g.,}\xspace}

\newcommand{\ie}{\emph{i.e.,}\xspace}
\newcommand{\Ie}{\emph{I.e.,}\xspace}

\newcommand{\true}{\mathsf{True}\xspace}
\newcommand{\True}{\textsf{True}\xspace}
\newcommand{\false}{\mathsf{False}\xspace}
\newcommand{\False}{\textsf{False}\xspace}
\newcommand{\sP}{\mathcal{P}\xspace}
\newcommand{\bZ}{{Z}\xspace} 

\newcommand{\capacity}{\mathsf{channelCapacity}\xspace} 
\newcommand{\done}{\mathsf{result}\xspace}
\newcommand{\bcdone}{\mathsf{result}\xspace}
\newcommand{\sameValue}{\mathit{sameValue}\xspace}
\newcommand{\binValues}{\mathit{binValues}\xspace}
\newcommand{\typ}{\emsA{\mathit{phs}\xspace}}
\newcommand{\Figure}{Fig.\xspace}
\usepackage{caption}

\newcommand{\respectively}{resp.\xspace}
\newcommand{\respectivelyC}{resp.,\xspace}
\newcommand{\respectivelyP}{resp.\xspace}

\newcommand{\B}{\vspace*{-\smallskipamount}}

\newenvironment{claimProof}{\par\noindent\textbf{Proof of Claim  \clmcnt\space}}{\hfill $\Box_{Claim ~ \clmcnt}$}

\newenvironment{lemmaProof}{\par\noindent\textbf{Proof of Lemma  \lemcnt\space}}{\hfill $\Box_{Lemma ~ \lemcnt}$}

\newenvironment{theoremProof}{\par\noindent\textbf{Proof of Theorem  \thmcnt\space}}{\hfill $\Box_{Theorem ~ \thmcnt}$}

\newcommand{\clmcnt}{0}
\newcommand{\lemcnt}{0}
\newcommand{\thmcnt}{0}

\newcommand{\Section}[1]{\section{#1}}
\newcommand{\Subsection}[1]{\subsection{#1}}
\newcommand{\Subsubsection}[1]{\subsubsection{#1}}

\usepackage{wrapfig}
\usepackage{url,fullpage}
\begin{document}
\captionsetup[figure]{name={Fig.}} 

\title{Self-stabilizing Byzantine- and Intrusion-tolerant Consensus\\\Large{(preliminary version)}}

		\author{Romaric Duvignau~\footnote{Department of Computer Science and Engineering, Chalmers University of Technology, Gothenburg, SE-412 96, Sweden. Email: \texttt{\{duvignau,elad\}@chalmers.se}} \and Michel Raynal~\footnote{IRISA, University Rennes 1, France and Polytechnic University, Hong Kong. Email: \texttt{michel.raynal@irisa.fr}} \and Elad M.\ Schiller}
%
%
%
%
%
	
\date{}
\maketitle

\begin{abstract}
	One of the most celebrated problems of fault-tolerant distributed computing is the consensus problem. It was shown to abstract a myriad of problems in which processes have to agree on a single value. Consensus applications include fundamental services for the environments of the Cloud or Blockchain. In such challenging environments, malicious behavior is often modeled as adversarial Byzantine faults. At OPODIS 2010, Most{\'{e}}faoui and Raynal, in short, MR, presented a Byzantine- and intrusion-tolerant solution to consensus in which the decided value cannot be a value proposed only by Byzantine processes. In addition to this validity property, MR has optimal resilience since it can deal with up to $t < n/3$ Byzantine processes, where $n$ is the number of processes. We note that MR provides this multivalued consensus object (which accepts proposals taken from a set with a finite number of values) assuming the availability of a single Binary consensus object (which accepts proposals taken from the set $\{0,1\}$).
	
	This work, which focuses on multivalued consensus, aims at the design of an even more robust solution than MR. Our proposal expands MR's fault-model with self-stabilization, a vigorous notion of fault-tolerance. In addition to tolerating Byzantine and communication failures, self-stabilizing systems can automatically recover after the occurrence of \emph{arbitrary transient-faults}. These faults represent any violation of the assumptions according to which the system was designed to operate (provided that the algorithm code remains intact).
	
	To the best of our knowledge, we propose the first self-stabilizing solution for intrusion-tolerant multivalued consensus for asynchronous message-passing systems prone to Byzantine failures. 
	%
\end{abstract}


\Section{Introduction}
\label{sec:intro}
%

\Subsection{Background and motivation} 
\label{sec:backgroundMotivation}
The consensus problem is one of the most challenging tasks in fault-tolerant distributed computing. The problem definition is rather simple. It assumes that each non-faulty process advocates for a single value from a given set $V$. The problem of \emph{Byzantine-tolerant Consensus} (BC) requires \emph{BC-completion}, \ie all non-faulty processes decide a value, \emph{BC-Agreement}, \ie no two non-faulty processes can decide different values, and \emph{BC-validity}, \ie if all non-faulty processes propose the same value $v\in V$, only $v$ can be decided. When the set, $V$, from which the proposed values are taken is $\{0,1\}$, the problem is called Binary consensus. Otherwise, it is named multivalued consensus. This work studies robust solutions to the problem of multivalued consensus that assume access to a single Binary consensus object. We aim at designing solutions that have higher degrees of dependability than the existing implementations.

\Subsection{Byzantine fault-tolerance} 
\label{sec:BYZnoIntro}
Lamport, Shostak, and Pease~\cite{DBLP:journals/toplas/LamportSP82} say that a process commits a Byzantine failure if it deviates from the algorithm instructions, say, by deferring (or omitting) messages that were sent by the algorithm or sending fake messages, which the algorithm never sent. Such malicious behavior can be the result of hardware malfunctions or software errors as well as coordinated malware attacks. In order to safeguard against such attacks, Most{\'{e}}faoui and Raynal~\cite{DBLP:conf/opodis/MostefaouiR10,DBLP:journals/acta/MostefaouiR17} as well as Correia, Neves, and Ver{\'{\i}}ssimo~\cite{DBLP:journals/cj/CorreiaNV06,DBLP:journals/tpds/NevesCV05} suggested the \emph{BC-no-intrusion} validity requirement (aka \emph{intrusion-tolerance}). Specifically, the decided value cannot be a value that was proposed only by faulty processes. Also, when it is not possible to decide on a value, the error symbol, $\blitza$, is returned.

For the sake of deterministic solvability~\cite{DBLP:journals/jacm/DworkLS88,DBLP:journals/toplas/LamportSP82,DBLP:journals/jacm/PeaseSL80,DBLP:conf/srds/Perry84}, we assume that there are at most $t<n/3$ Byzantine processes in the system, where $n$ is the total number of processes. It is also well-known that no deterministic (multivalued or Binary) consensus solution exists for asynchronous systems in which at least one process may crash (or one process can be Byzantine)~\cite{DBLP:journals/jacm/FischerLP85}. The studied multivalued consensus algorithms circumvent this impossibility by assuming that the system model is enriched with a Byzantine-tolerant object that solves Binary consensus. 
This is as in the studied solution by Most{\'{e}}faoui and Raynal~\cite{DBLP:conf/opodis/MostefaouiR10}, MR from now on, \ie reducing multivalued consensus to Binary consensus.

\Subsection{Self-stabilization} 
We study an asynchronous message-passing system that has no guarantees on the communication delay and the algorithm cannot explicitly access the local clock. Our fault model includes undetectable Byzantine failures.
In addition to the failures captured by our model, we also aim to recover from \emph{arbitrary transient-faults}, \ie any temporary violation of assumptions according to which the system and network were designed to operate. 
This includes the corruption of control variables, such as the program counter and message payloads, as well as operational assumptions, such as that at most $t<n/3$ processes are not faulty. 
Since the occurrence of these failures can be arbitrarily combined, we assume that these transient-faults can alter the system state in unpredictable ways. 
In particular, when modeling the system, Dijkstra~\cite{DBLP:journals/cacm/Dijkstra74} assumes that these violations bring the system to an arbitrary state from which a \emph{self-stabilizing system} should recover, see~\cite{DBLP:series/synthesis/2019Altisen,DBLP:books/mit/Dolev2000} for details. 
\Ie Dijkstra requires (i) recovery after the last occurrence of a transient-fault and (ii) once the system has recovered, it must never violate the task requirements. 
Arora and Gouda~\cite{DBLP:journals/tse/AroraG93} refer to the former requirement as the \emph{Closure} property and to the latter requirement as the \emph{Convergence} property.

\Subsection{Related work}
Ever since the seminal work of Lamport, Shostak, and Pease~\cite{DBLP:journals/toplas/LamportSP82} four decades ago, Byzantine \emsA{fault-tolerant (BFT)} consensus has been an active research subject, see~\cite{DBLP:journals/ijccbs/CorreiaVNV11}. The recent rise of distributed ledger technologies, \eg~\cite{DBLP:conf/sp/AbrahamMN0Y20}, brought phenomenal attention to the subject since Blockchain technology market worth is expected to reach $395$ B USD by 2028.~\footnote{\url{www.grandviewresearch.com/industry-analysis/blockchain-technology-market}.} Therefore, we aim to provide a degree of dependability that is higher than existing solutions.  

Ben-Or, Kelmer, and Rabin~\cite{DBLP:conf/podc/Ben-OrKR94} presented the first reduction from \emsA{BFT} multivalued consensus to \emsA{BFT} Binary consensus. 
They do not consider intrusion tolerance. 
As mentioned, Most{\'{e}}faoui and Raynal~\cite{DBLP:conf/opodis/MostefaouiR10,DBLP:journals/acta/MostefaouiR17} as well as Correia, Neves, and Ver{\'{\i}}ssimo~\cite{DBLP:journals/cj/CorreiaNV06,DBLP:journals/tpds/NevesCV05} proposed the notion of intrusion tolerance. 
Our contribution is a self-stabilizing variation on MR. In other words, we offer an algorithm for multivalued consensus that is self-stabilizing BFT, in short SSBFT.

\begin{figure}
	\begin{center}
		\includegraphics[scale=0.35, clip]{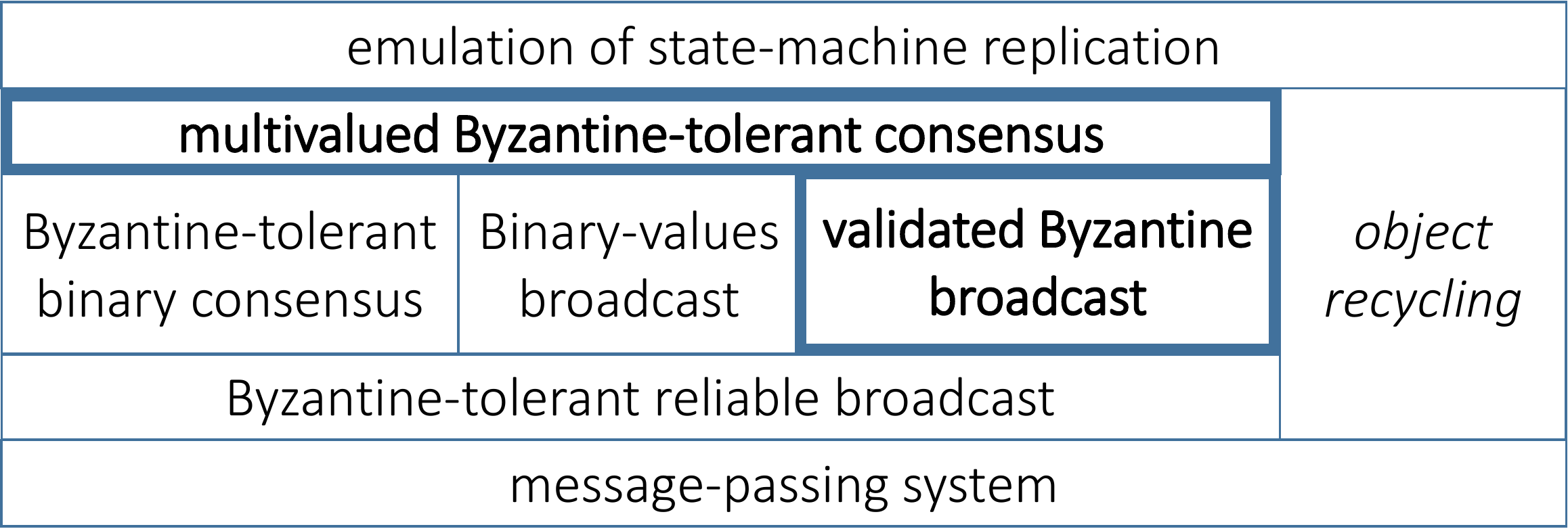}
	\end{center}
	\caption{\label{fig:suit}{The studied architecture assumes the availability of an SSBFT protocol for Binary consensus and an SSBFT mechanism for object recycling. The studied problems appear in boldface fonts. The other layers mentioned in the text above are in plain font, \ie SSBFT BRB, SSBFT BV-broadcast, and SSBFT state machine emulation.}}
\end{figure}

There are (non-self-stabilizing) BFT solutions~\cite{DBLP:books/sp/Raynal18} and (crash-tolerant) self-stabilizing solutions~\cite{DBLP:conf/netys/BlanchardDBD14,DBLP:journals/jcss/DolevKS10,DBLP:conf/edcc/LundstromRS21,DBLP:conf/icdcn/LundstromRS21}.   
Most{\'{e}}faoui, Moumen, and Raynal~\cite{DBLP:journals/jacm/MostefaouiMR15,DBLP:conf/podc/MostefaouiMR14} presented BFT algorithms for solving Binary consensus using common coins. Recently, Georgiou \etal \cite{DBLP:journals/corr/abs-2103-14649} proposed a self-stabilizing variation on the one in \cite{DBLP:conf/podc/MostefaouiMR14} that satisfies the safety requirements, \ie agreement and validity, with an exponentially high probability that depends only on a predefined constant. Georgiou \etal's solution can be used as the SSBFT Binary consensus object needed for our solution.    

The most related work to our includes SSBFT state-machine replication by Binun \etal~\cite{DBLP:conf/sss/BinunCDKLPYY16,DBLP:conf/cscml/BinunDH19} for synchronous systems and Dolev \etal~\cite{DBLP:conf/cscml/DolevGMS18} for practically-self-stabilizing partially-synchronous systems.
Note that both Binun \etal and Dolev \etal study another problem for another kind of system settings. 
In the broader context of SSBFT solutions for message-passing systems, the literature studied extensively the problems of clock synchronization~\cite{DBLP:journals/corr/abs-2203-14016,perner2013byzantine,DBLP:conf/sss/Malekpour06,DBLP:conf/wdag/DolevH07,DBLP:conf/icpads/YuZY21,DBLP:conf/podc/DaliotDP04,DBLP:conf/sss/DolevH07,DBLP:conf/podc/Ben-OrDH08,DBLP:conf/sss/HochDD06,DBLP:conf/podc/DolevW95,DBLP:journals/jacm/LenzenR19,DBLP:journals/mst/KhanchandaniL19}, storage~\cite{DBLP:journals/tcs/BonomiPP18,DBLP:conf/icdcn/BonomiPP16,DBLP:conf/sss/BonomiPPT18,DBLP:conf/srds/BonomiPPT17,DBLP:conf/podc/BonomiPPT16,DBLP:conf/ipps/BonomiPT15,DBLP:conf/podc/BonomiDPR15}, and gathering of mobile robots~\cite{DBLP:conf/sss/AshkenaziDKKOW21,DBLP:conf/ic-nc/AshkenaziDKOW19,DBLP:journals/corr/DefagoP0MPP16,DBLP:journals/dc/DefagoPP20}. 
We also find solutions for link-coloring~\cite{DBLP:conf/opodis/MasuzawaT05,DBLP:conf/opodis/SakuraiOM04}, topology discovery~\cite{DBLP:conf/netys/DolevLS13,DBLP:journals/tpds/NesterenkoT09}, overlay networks~\cite{DBLP:conf/opodis/DolevHR07},  exact agreement~\cite{DBLP:conf/podc/DaliotD06} approximate agreement~\cite{DBLP:journals/tcs/BonomiPPT19}, asynchronous unison~\cite{DBLP:journals/jpdc/DuboisPNT12}, communication in dynamic networks~\cite{DBLP:conf/opodis/Maurer20}, and reliable broadcast~\cite{DBLP:journals/corr/abs-2201-12880,DBLP:conf/srds/MaurerT14}. 

\Subsection{Demonstrating self-stabilization in the studied architecture}
\label{sec:arch}    
Many Cloud computing and distributed ledger technologies are based on state-machine replication. 
Following Raynal~\cite[Ch. 16 and 19]{DBLP:books/sp/Raynal18}, \Figure~\ref{fig:suit} illustrates how total order broadcast can facilitate the ordering of the automaton's state transitions. 
This order can be defined by instances of multivalued consensus objects, which in turn, invokes Binary consensus \emsA{and Binary-values broadcast (in short BV-broadcast),} such as the SSBFT one by Georgiou \etal~\cite{DBLP:journals/corr/abs-2103-14649} as well as \emph{Byzantine-tolerant Reliable Broadcast} (in short BRB), such as the SSBFT solution by Duvignau, Raynal, and Schiller~\cite{DBLP:journals/corr/abs-2201-12880}.
This work focuses on transforming the non-self-stabilizing MR solution for Byzantine- and intrusion-tolerant multivalued consensus into one that is self-stabilizing and Byzantine- and intrusion-tolerant.

Just as MR, we do not focus on the management of consensus invocations since we assume the availability of a mechanism for eventually recycling all consensus objects that have completed their tasks. 
Georgiou \etal use such mechanisms in~\cite{DBLP:conf/netys/GeorgiouMRS21}. 
In their extended version~\cite{DBLP:journals/corr/abs-2103-14649}, they detail the mechanism construction.

When using only a predefined number of objects, the availability of the SSBFT recycling mechanism allows for the devising of an elegant solution that is based on a code transformation of the non-self-stabilizing BFT MR algorithm to an SSBFT one. The transformation concentrates on assuring operation completion since once all objects have been recycled, the system reaches its \emph{post-recycling state}, which has no remanence of stale information. Thus, starting at this state, the system behavior is similar to the one of the non-self-stabilizing BFT MR algorithm.

As mentioned, transient faults are modeled to leave the system in an arbitrary state. In order to guarantee the operation completion when starting in an arbitrary state, we identify proof invariants that their violation (due to state corruption) can prevent operation completion. Based on these invariants, we transform the non-self-stabilizing BFT MR algorithm into an SSBFT one via the inclusion of invariant tests.

Our correctness proof demonstrates recovery after the occurrence of the last transient fault by showing that any operation, using the added invariant tests, eventually returns a value that indicates operation completion. In other words, we demonstrate that when starting in an arbitrary system state, eventually, all objects become recyclable. As explained above, by eventually recycling all of these objects, the system arrives at a post-recycling state. For the sake of completeness, our proof also shows that, starting at a post-recycling state, the system satisfies the task requirements, which is multivalued consensus.

We clarify that we do not deviate from the analytical framework proposed by Arora and Gouda~\cite{DBLP:journals/tse/AroraG93}, which requires the demonstration of the Closure and the Convergence properties. As mentioned, our correctness proof demonstrates Convergence by showing that the components used and proposed by our solution always eventually become recyclable. Once they are all recycled, the system is in its post-recycling state. Starting from that state, Closure is proved.

\Subsection{Our contribution}
We present a fundamental module for dependable distributed systems: an SSBFT algorithm for multivalued consensus for asynchronous message-passing systems. 
We obtain this new self-stabilizing algorithm via a transformation of the non-self-stabilizing MR algorithm by Most{\'{e}}faoui and Raynal~\cite{DBLP:conf/opodis/MostefaouiR10}. 
MR offers optimal resilience by assuming $t < n/3$, where $t$ is the number of faulty processes and $n$ is the total number of processes. The proposed solution preserves this optimality. 

In the absence of transient-faults, our solution achieves consensus within a constant time as in the MR algorithm. 
After the occurrence of any finite number of arbitrary transient-faults, the system recovers eventually.
The communication costs of the studied and proposed algorithms are similar in the number of BRB and Binary consensus invocations. \emsA{The main difference is that our SSBFT solution uses BV-broadcast for making sure that the value decided by the SSBFT Binary consensus object remains consistent until the proposed SSBFT solution completes its task and is ready to be recycled.}

To the best of our knowledge, we propose the first self-stabilizing Byzantine- and intrusion-tolerant algorithm for solving multivalued consensus in asynchronous message-passing systems that are enriched by a single SSBFT Binary consensus object and two SSBFT BRB objects. 
We believe that our solution can stimulate research for the design of algorithms for the environments of the Cloud and distributed ledger technologies that are far more robust than the existing implementations since the latter cannot recover after the occurrence of transient faults.

\medskip

For the reader's convenience, Table~\ref{fig:Glossary} in the Appendix includes the Glossary, where all abbreviations are listed. 

\Section{System Settings}
\label{sec:sys} 
We consider an asynchronous message-passing system that has no guarantees on the communication delay. Also, the algorithm cannot explicitly access the (local) clock (or use timeout mechanisms). The system consists of a set, $\sP$, of $n$ fail-prone nodes (or processes) with unique identifiers. Any pair of nodes $p_i,p_j \in \sP$ has access to a bidirectional communication channel, $\mathit{channel}_{j,i}$, that, at any time, has at most $\capacity \in \bZ^+$ packets on transit from $p_j$ to $p_i$ (this assumption is due to a known impossibility~\cite[Chapter 3.2]{DBLP:books/mit/Dolev2000}).

%
\label{sec:interModel}
In the \emph{interleaving model}~\cite{DBLP:books/mit/Dolev2000}, the node's program is a sequence of \emph{(atomic) steps}. Each step starts with an internal computation and finishes with a single communication operation, \ie a message $send$ or $receive$. The \emph{state}, $s_i$, of node $p_i \in \sP$ includes all of $p_i$'s variables and $\mathit{channel}_{j,i}$. The term \emph{system state} (or configuration) refers to the tuple $c = (s_1, s_2, \cdots,  s_n)$. We define an \emph{execution (or run)} $R={c[0],a[0],c[1],a[1],\ldots}$ as an alternating sequence of system states $c[x]$ and steps $a[x]$, such that each $c[x+1]$, except for the starting one, $c[0]$, is obtained from $c[x]$ by $a[x]$'s execution.


\Subsection{The fault model and self-stabilization}
The \emph{legal executions} ($LE$) set refers to all the executions in which the requirements of task $T$ hold. In this work, $T_{\text{MVC}}$ denotes the task of multivalued consensus, which Section~\ref{sec:intro} specifies, and the executions in the set $LE_{\text{MVC}}$ fulfill $T_{\text{MVC}}$'s requirements.

%
%



\Subsubsection{Arbitrary node failures.}
\label{sec:arbitraryNodeFaults}
Byzantine faults model any fault in a node including crashes, and arbitrary malicious behaviors. Here the adversary lets each node receive the arriving messages and calculate its state according to the algorithm. However, once a node (that is captured by the adversary) sends a message, the adversary can modify the message in any way, delay it for an arbitrarily long period or even remove it from the communication channel. The adversary can also send messages spontaneously. Note that the adversary has the power to coordinate such actions without any limitation on his computational or communication power. 
%
%
For the sake of solvability~\cite{DBLP:journals/toplas/LamportSP82,DBLP:journals/jacm/PeaseSL80,DBLP:conf/podc/Toueg84}, the fault model that we consider limits only the number of nodes that can be captured by the adversary. That is, the number, $t$, of Byzantine failures needs to be less than one-third of the number, $n$, of nodes in the system, \ie $3t+1\leq n$. The set of non-faulty nodes is denoted by $\Correct$ and called the set of correct nodes.

\Subsubsection{Arbitrary transient-faults}
\label{sec:arbitraryTransientFaults}
We consider any temporary violation of the assumptions according to which the system was designed to operate. We refer to these violations and deviations as \emph{arbitrary transient-faults} and assume that they can corrupt the system state arbitrarily (while keeping the program code intact). The occurrence of a transient fault is rare. Thus, we assume that the last arbitrary transient fault occurs before the system execution starts~\cite{DBLP:books/mit/Dolev2000}. Also, it leaves the system to start in an arbitrary state.

\Subsection{Dijkstra's self-stabilization}
\label{sec:Dijkstra}
An algorithm is \emph{self-stabilizing} with respect to $LE$, when every execution $R$ of the algorithm reaches within a finite period a suffix $R_{legal} \in LE$ that is legal. Namely, Dijkstra~\cite{DBLP:journals/cacm/Dijkstra74} requires $\forall R:\exists R': R=R' \circ R_{legal} \land R_{legal} \in LE \land |R'| \in \bZ^+$, where the operator $\circ$ denotes that $R=R' \circ R''$ is the concatenation of $R'$ with $R''$. 
The part of the proof that shows the existence of $R'$ is called the \emph{convergence} (or recovery) proof, and the part that shows that $R_{legal} \in LE$ is called the \emph{closure} proof. 
Recall that in Section~\ref{sec:arch}, we explain the connection between convergence and closure as well as the SSBFT recycling mechanism, SSBFT recyclable objects, and the post-recycling state.


\Subsection{External Building blocks}
\ems{As mentioned, we assume the availability of an SSBFT recycling mechanism (Section~\ref{sec:arch}) for BRB and consensus objects, as the ones proposed by Georgiou \etal~\cite{DBLP:conf/netys/GeorgiouMRS21,DBLP:journals/corr/abs-2103-14649} and Duvignau, Raynal, and Schiller~\cite{DBLP:journals/corr/abs-2201-12880}. Recycling occurs eventually after all of the objects complete their tasks. We specify below the object that this paper assumes to be available.}

\Subsubsection{SSBFT Byzantine-tolerant Reliable Broadcast (BRB)}
\label{sec:BRBext}
\emsA{The communication abstraction of Byzantine Reliable Broadcast (BRB) allows every node to invoke the $\mathsf{broadcast}(v):v\in V$ and $\mathsf{deliver}(k):p_k \in \sP$ operations.}



\begin{definition}
	\label{def:prbDef}
	The operations $\mathsf{broadcast}(v)$ and $\mathsf{deliver}(k)$ should satisfy the following.
	\begin{itemize}
		\item \textbf{BRB-validity.~~} Suppose a correct node BRB-delivers message $m$ from a correct node $p_i$. Then, $p_i$ had BRB-broadcast $m$.
		\item \textbf{BRB-integrity.~~} No correct node BRB-delivers more than once.
		\item \textbf{BRB-no-duplicity.~~} No two correct nodes BRB-deliver different messages from $p_i$ (who might be faulty).
		\item \textbf{BRB-completion-1.~~} Suppose $p_i$ is a correct sender. All correct nodes BRB-deliver from $p_i$ eventually.
		\item \textbf{BRB-completion-2.~~} Suppose a correct node BRB-delivers a message from $p_i$ (who might be faulty). All correct nodes BRB-deliver $p_i$'s message eventually.
	\end{itemize}
\end{definition}

\emsA{We assume the availability of an SSBFT BRB implementation, such as the one by Duvignau, Raynal, and Schiller~\cite{DBLP:journals/corr/abs-2201-12880}. Such implementation lets} \ems{$p_i \in \sP$ to use the operation $\mathsf{deliver}_i(k)$ for retrieving the current return value, $v$, of the BRB broadcast from $p_k \in \sP$. 
	Before the completion of the task of the $\mathsf{deliver}_i(k)$ operation, $v$'s value is $\bot$. 
	This way, whenever $\mathsf{deliver}_i(k)\neq \bot$, node $p_i$ knows that the task is completed and the returned value can be used.}


\Subsubsection{SSBFT Binary-values Broadcast (BV)}
\label{sec:sefBVbrodcast}
\emsA{This is an all-to-all broadcast operation of Binary values. It uses the operation, $\mathsf{bvBroadcast}(v)$, which is assumed to be invoked by all the correct nodes, where $v,w \in \{0,1\}$. The set of values that are BV-delivered to node $p_i$ can be retrieved via the function $\binValues_i()$, which returns $\emptyset$ before the arrival of any $\mathsf{bvBroadcast}()$ by a correct node. We specify under which conditions values are added to $\binValues()$.} 

\begin{itemize}
	
	\item  \emsA{\textbf{BV-validity.} Suppose that $v \in \binValues_i()$ and $p_i$ is correct. It holds that $v$ has been BV-broadcast by a correct node.}
	
	\item \emsA{\textbf{BV-uniformity.} $v \in \binValues_i()$ and $p_i$ is correct. Eventually $\forall j \in \Correct: v \in \binValues_j()$.}
	
	\item \emsA{\textbf{BV-completion.} Eventually $\forall i \in \Correct: \binValues_i() \neq \emptyset$ holds.}
	
\end{itemize}

\emsA{The above requirements imply that eventually $\exists s \subseteq \{0,1\}: s \neq \emptyset \land \forall i \in \Correct: \binValues_i()=s$ and the set $s$ does not include values that were BV-broadcast only by Byzantine nodes. We note the existing SSBFT solutions for BV-broadcast Georgiou \etal~\cite{DBLP:journals/corr/abs-2103-14649}, which we use. Georgiou \etal's implementation allows the correct nodes to repeat a BV-broadcast using the same BV object. Our proof uses the fact that, as long as the correct nodes do not change their BV-broadcast messages, the requirements above hold.}


\Subsubsection{SSBFT Binary Consensus}
\ems{As mentioned, the studied solution reduces multivalued consensus to Binary consensus by enriching the system model with a BFT object that solves Binary consensus} \emsA{(Definition~\ref{def:consensus}).} 
\begin{definition}
	\label{def:consensus}
	\emsA{Every $p_i\in \sP$ has to propose a value $v_i \in V=\{\False,\True\}$ via an invocation of the $\mathsf{propose}_i(v_i)$ operation. (We prefer $V=\{\False,\True\}$ over the traditional representation of $V= \{0,1\}$.) Let $\mathit{Alg}$ be an algorithm that solves Binary  consensus. $\mathit{Alg}$ has to satisfy \emph{safety}, \ie BC-validity and BC-agreement, and \emph{liveness}, \ie BC-completion, requirements.}
	\begin{itemize}
		\item \emsA{\textbf{BC-validity.~~} 
			The value $v \in \{\False,\True\}$ decided by a correct node is a value proposed by a correct node.}
		\item \emsA{\textbf{BC-agreement.~~} Any two correct nodes that decide, do so with identical decided values.}
		\item \emsA{\textbf{BC-completion.~~} All correct nodes decide.}
	\end{itemize}
\end{definition}

\emsA{We assume that availability of SSBFT Binary consensus, such as Georgiou \etal~\cite{DBLP:journals/corr/abs-2103-14649}, which might fail to decide with negligible probability.} 
\ems{If that failure occurs, Georgiou \etal's solution might return the error symbol, $\blitza$, instead of a legitimate value from the set $\{0,1\}$.}


\Section{The Studied Algorithms}
\ems{The MR solution is based on a reduction of the studied problem to BFT Binary consensus. 
MR guarantees that the decided value is not a value proposed only by Byzantine nodes. 
Also, if there is a value, $v \in V$, that all correct nodes propose, then $v$ is decided. 
Otherwise, the decided value is either a value proposed by the correct nodes or the error symbol, $\blitza$. 
This way, an adversary that command its captured nodes to propose the same value, say, $v_{byz} \in V$, cannot lead to the selection of $v_{byz}$ without the support of at least one correct node. 
As depicted in \Figure~\ref{fig:suit}, the MR reduction is based on a communication abstraction, named Validated Byzantine Broadcast, in short VBB, which we present in Section~\ref{sec:VBB} before the reduction itself (Section~\ref{sec:MVC}).}


\Subsection{Validated Byzantine Broadcast (VBB)} 
\label{sec:VBB}
This communication abstraction sends messages from all nodes to all nodes. 
%
%
It offers the operation, $\mathsf{vbbBroadcast}(v)$ and raises the event $\mathsf{vbbDeliver}(d)$, for VBB-broadcasting, and \respectivelyC VBB-delivering messages.

\Subsubsection{\ems{Specifications}} 
\label{sec:specVVB}
We detail VBB-broadcast requirements below.

\begin{itemize}
\item \textbf{VBB-validity.~~} VBB-delivery of messages needs to relate to VBB-broadcast of messages in the following manner.
\begin{itemize}
	\item \textbf{VBB-justification.~~} Suppose $p_i : i \in \Correct$ VBB-delivers message $m\neq \blitza$ from some (faulty or correct) node. There is at least one correct node that VBB-broadcast $m$.
	
	\item \textbf{VBB-obligation.} Suppose all correct nodes VBB-broadcast the same $v$. All correct nodes VBB-delivers $v$ from each correct node.
\end{itemize}

\item \textbf{VBB-uniformity.~~}  Let $p_i:i \in \Correct$. Suppose VBB-delivers $m' \in \{m,\blitza\}$ from a (possibly faulty) node $p_j$. All the correct nodes VBB-deliver the same message $m'$ from $p_j$.

\item \textbf{VBB-completion.~~} Suppose $p_i$ VBB-broadcasts $m$, such that $i \in \Correct$. All the correct nodes VBB-deliver from $p_i$.
\end{itemize}

\ems{We also say that a} complete VBB-broadcast instance includes $\mathsf{vbbBroadcast}_i(m_i)$ invocation by every correct node $p_i \in \sP$.
It also includes $\mathsf{vbbDeliver}()$ of $m'$ from at least $(n\mathit{-}t)$ distinct nodes, where $m'$ is either $p_j$'s message, $m_j$, or the error symbol, $\blitza$. 
The latter value is returned when a message from a given sender cannot be validated. 
This validation requires $m_j$ to be VBB-broadcast by at least one correct node. 
That is, to be VBB delivered from at least $(t\mathit{+}1)$ different nodes (including its sender $p_j$), because no node $p_i$ can foresee its prospective failures, \eg due to unexpected crashes. 

\begin{algorithm}[t!]
\begin{\algSize}	
	
	\medskip
	
	\textbf{operation} $\mathsf{vbbBroadcast}(v)$ \label{ln:vbbBradcastAAA} \Begin{
		
		BRB-broadcast $\mathrm{INIT}(i, v)$\label{ln:brbBradcast0}\; 
		
		\textbf{wait} $|rec|$$\geq$$n\mathit{-}t$ \textbf{where} $rec$ is the multiset of BRB-delivered values\label{ln:brbBradcast0wait}\;
		
		
		BRB-broadcast $\mathrm{VALID}(i, (\mathit{equal}(v, rec) \geq  n \mathit{-} 2t))$\label{ln:brbBradcast1}\; 
		
	}
	
	\smallskip
	
		
		\ForEach{$p_j \in \sP$ \emph{execute concurrently}\label{ln:vbbBackground}}{
			
			\textbf{wait} $\mathrm{INIT}(j,v)$ and $\mathrm{VALID}(j,x)$ BRB-delivered from $p_j$\label{ln:vbbWaitValidINIT}\;
			
			\lIf{$x$\label{ln:ifXtrue}}{\{\textbf{wait} $(\mathit{equal}(v, rec) \geq n \mathit{-} 2t)$; $d \gets v$\}} 
			\lElse{\{\textbf{wait} $(\mathit{differ}(v, rec) \geq t \mathit{+} 1)$; $d \gets \blitza$\label{ln:ifXtrueElse}\}}
			
			$\mathsf{vbbDeliver}(d)$ at $p_i$ as the value VBB-broadcast by $p_j$\label{ln:vbbDeliverA}\;
			

	}
	
	\medskip
	
	\caption{\label{alg:vbbBroadcast}Non-self-stabilizing BFT VBB-broadcast; code for $p_i$}
\end{\algSize}
\end{algorithm}

\Subsubsection{Implementing VBB-broadcast}
Algorithm~\ref{alg:vbbBroadcast} presents the studied VBB-broadcast. 


\noindent \textbf{Notation:~~} 
Let $|rec|$ denote the number of elements in the multiset $rec$. We use\reduce{ the functions} $\mathit{equal}(v, rec)$ and $\mathit{differ}(v, rec)$ to return the number of occurrences in $rec$ that are equal to, and \respectivelyC different from $v$. 

\noindent \textbf{Overview:~~} 
Algorithm~\ref{alg:vbbBroadcast} invokes BRB-broadcast twice in the first part of the algorithm (lines~\ref{ln:vbbBradcastAAA} to~\ref{ln:brbBradcast1}) and then VBB-delivers messages from nodes in the second part (lines~\ref{ln:vbbBackground} to~\ref{ln:vbbDeliverA}).

Node $p_i$ first BRB-broadcasts $\mathrm{INIT}(i, v_i)$ (where $v_i$ is the VBB-broadcast message), and suspends until the arrival of $\mathrm{INIT}()$ from at least $(n \mathit{-} t)$ different nodes (lines~\ref{ln:brbBradcast0} to~\ref{ln:brbBradcast0wait}), which $p_i$ collects in the multiset $rec_i$. In line~\ref{ln:brbBradcast0}, node $p_i$ tests whether $v_i$ was  BRB-delivered from at least $n\mathit{-}2t \geq t\mathit{+}1$ different nodes. Since this means that $v_i$ was BRB-broadcast by at least one correct node, $p_i$ attests to the validity of $v_i$ (line~\ref{ln:brbBradcast1}). Recall that each time $\mathrm{INIT}()$ arrives at $p_i$, the message is added to $rec_i$. Therefore, the fact that $|rec_i| \geq n \mathit{-} t$ holds (line~\ref{ln:brbBradcast0wait}) does not keep $rec_i$ from growing.

Algorithm~\ref{alg:vbbBroadcast}'s second part (lines~\ref{ln:vbbBackground} to~\ref{ln:vbbDeliverA}) includes $n$ concurrent background tasks. Each task aims at VBB-delivering a message from a different node, say, $p_j$. It starts by waiting until $p_i$ BRB-delivered both $\mathrm{INIT}(j, v_j)$ and $\mathrm{VALID}(j, x_j)$ from $p_j$ so that $p_i$ has both $p_j$'s VBB's values, $v_j$, and the result of its validation test, $x_j$. 

\begin{itemize}
\item \textbf{The case of $x_j=\true$ (line~\ref{ln:ifXtrue}).~~} Since $p_j$ might be faulty, we cannot be sure that $v_j$ was indeed validated. Thus, $p_i$ re-attests $v_j$ by waiting until $\mathit{equal}(v_j, rec_i) \geq n\mathit{-}2t$ holds. If this ever happens, $p_i$ VBB-delivers $v_j$ as a message from $p_j$, because the wait condition implies that $\mathit{equal}(v_j, rec_i) \geq t \mathit{+} 1$ since $n\mathit{-}2t \geq t\mathit{+}1$.

\item \textbf{The case of $x_j=\false$ (line~\ref{ln:ifXtrueElse}).~~} For similar reasons to the former case, $p_i$ needs to wait until $rec_i$ contains at least $t\mathit{+}1$ items that are not $v_j$, because this implies that at least one correct note cannot attest $v_j$'s validity. If this ever happens, $p_i$ VBB-delivers the error symbol, $\blitza$, as the received message from $p_j$.
\end{itemize}



\Subsubsection{\emsA{Invariants that could be violated due to transient faults}}
\label{sec:vbbInv}
\emsA{The occurrence of a transient fault can violate the following invariants, which an SSBFT solution needs to address.}

\begin{enumerate}
\item\label{itm:ValidWithoutInit} \emsA{The state of node $p_i$ must not encode the occurrence of BRB execution of phase $\valid$ (line~\ref{ln:brbBradcast1}) without encoding BRB execution of phase $\init$ (line~\ref{ln:brbBradcast0}).} 

\item\label{itm:badFormat} \emsA{For a given phase, $\typ\in\texttt{vbbMSG}$, the format of a message that is BRB-delivered must follow the one of BRB-broadcast of phase $\typ$, \ie $(k,v):p_k \in \sP \land v \in V$ for phase $\init$ and $(k,x):p_k \in \sP \land x \in \{\false,\true\}$ for phase $\valid$.} 

\item\label{itm:nextPhase} \emsA{For a given phase, $\typ\in\texttt{vbbMSG}$, if at least $n-t$ different nodes BRB-delivered messages of phase $\typ$, to node $p_i$, the state of $p_i$ must lead to the next phase, \ie from $\init$ to $\valid$, or from $\valid$ to operation complete, in which VBB-deliver a non-$\bot$ value.} 	
\end{enumerate}

\begin{algorithm}[t!]
\begin{\algSize}

	\medskip
	
	\textbf{variables:}
	$\mathit{bcO}:=\bot$; \texttt{/*} Binary consensus object, $\bot$ is the initial state. \texttt{*/};
	
	\medskip
	
	\textbf{macro} $\sameValue()$ \textbf{do return} $\exists v$ $\neq$ $\blitza:\mathit{equal}(v, rec)$ $\geq$ $ n \mathit{-} 2t \land rec=\{v'$ $\neq$ $\blitza\}$\label{ln:mvcLet} \textbf{where} $rec$ is a multiset of the\remove{ $(n\mathit{-}t)$} values VBB-delivered (line~\ref{ln:mvcWait})
	
	\medskip
	
	\textbf{operation} $\mathsf{propose}(v)$ \label{ln:mvcPropuse} \Begin{
		
		$\mathsf{vbbBroadcast}$ $\mathrm{EST}(v)$\label{ln:mvcESTsend}\; 
		
		\textbf{wait} $\mathrm{EST}(\bullet)$ messages VBB-delivered from $(n\mathit{-}t)$ different nodes\label{ln:mvcWait}\;
		
		
		\lIf{$\neg bcO.\mathsf{propose}(\sameValue())$\label{ln:mvcIf}}{\Return{$\blitza$}\label{ln:mvcThen}}
		
		\lElse{\textbf{wait} $(\exists v\neq \bot:\mathit{equal}(v,rec)\geq n \mathit{-}2t)$ \Return($v$)\label{ln:mvcElse}}
		
	}

	\medskip
	
	\caption{\label{alg:reductionNon}Non-self-stabilizing BFT multivalued consensus; code for $p_i$}
\end{\algSize}
\end{algorithm}

\Subsection{Multivalued Byzantine-tolerant Consensus}
\label{sec:MVC}
Algorithm~\ref{alg:reductionNon} reduces any instance of the BFT multivalued consensus problem to BFT Binary consensus in message-passing systems that have up to $t < n/3$ Byzantine nodes.
\emsA{Algorithm~\ref{alg:reductionNon} uses VBB-broadcast abstraction (Algorithm~\ref{alg:vbbBroadcast}).} 
Note that the line numbers of Algorithm~\ref{alg:reductionNon} continue the ones of Algorithm~\ref{alg:vbbBroadcast}. 

\Subsubsection{\ems{Specifications}} 
Recall the task of multivalued Byzantine- and intrusion-tolerant consensus includes the requirements of BC-validity, BC-agreement, and BC-completion (Section~\ref{sec:backgroundMotivation}) as well as the BC-no-Intrusion property (Section~\ref{sec:BYZnoIntro}). 

\Subsubsection{\ems{Implementation}} 
%
%
%
Node $p_i$ has to wait for $\mathrm{EST}()$ messages from $(n \mathit{-} t)$ different nodes after it as VBB-broadcast its own value (lines~\ref{ln:mvcESTsend} to~\ref{ln:mvcWait}). It holds all the VBB-delivered values in the multiset $rec_i$ (line~\ref{ln:mvcLet}) before testing whether $rec_i$ includes (1) non-$\blitza$ replies from  at least $(n \mathit{-} 2t)$ different nodes, and (2) exactly one non-$\blitza$ value $v$ (line~\ref{ln:mvcLet}). The test result is proposed to the Binary consensus object, $\mathit{bcO}$ (line~\ref{ln:mvcIf}).

Once consensus was reached, $p_i$ decides according to the consensus result, $\mathit{bcO}_i.\bcdone()$. Specifically, if $\mathit{bcO}_i.\bcdone() =\false$, $p_i$ returns the error symbol, $\blitza$, since there is no guarantee that any correct node was able to attest to the validity of the proposed value. Otherwise, $p_i$ waits until it received $\mathrm{EST}(v)$ messages that have identical values from at least $(n \mathit{-} 2t)$ different nodes (line~\ref{ln:mvcElse}) before returning that value $v$. Note that some of these $(n \mathit{-} 2t)$ messages were already VBB-delivered at line~\ref{ln:mvcWait}. The proof in~\cite{DBLP:conf/opodis/MostefaouiR10} shows that any correct node that invokes $\mathit{bcO}_i.\mathsf{propose}(\true)$ does so if all correct nodes eventually VBB-deliver identical values at least $(n \mathit{-} 2t)$ times. Then, any correct node can decide on the returned value for the multivalued consensus object once it also VBB-delivers identical values at least $(n \mathit{-} 2t)$ times.

\Subsubsection{\emsA{Invariants that could be violated due to transient faults}}
\label{sec:mvcInv}
\emsA{The occurrence of a transient fault can let the Binary consensus object decide on a value never proposed, \ie violates BC-validity. 
Any SSBFT solution needs to address this concern since the multivalued consensus object can block indefinitely if $bcO$ decides $\True$ when for all correct nodes, $p_i$, $\sameValue_i()=\False$ holds.}

%

\begin{algorithm*}[t!]
\begin{\algSize}

	\medskip
	
	\textbf{types:}
	\label{ln:types} \remove{$\texttt{brbMSG} :=\{\init, \echo, \ready\}$; } $\texttt{vbbMSG} :=\{\init, \valid\}$; 
	%
	%
	
	\medskip
	
	\textbf{variables:}
	$\mathit{brb}[\texttt{vbbMSG}][\sP]:=[[\bot,\ldots,\bot],[\bot,\ldots,\bot]]$ \texttt{/*} Two \emsA{phases} of BRB objects. The value $\bot$ represents the post-recycling state. \texttt{*/}

	
	

\medskip

\textbf{macros:}
$\mathit{vbbEcho}(\typ)$ \label{ln:auxVSsP2tpkS}\textbf{do return} $\exists_{S \subseteq\sP: n \mathit{-}t\leq |S|} \forall_{p_k \in S} \, \mathit{brb}[\typ][k].\mathsf{deliver}() \neq \bot$\;

$\mathit{vbbEq}(\typ,v) := \exists_{S \subseteq\sP: n \mathit{-}2t\leq |S|} \forall_{p_\ell \in S} (v = \mathit{brb}[\typ].\mathsf{deliver}(\ell))$\label{ln:vbbEq}\; 

$\mathit{vbbDiff}(\typ,v) :=\exists_{S \subseteq\sP: t\mathit{+}1\leq |S|} \forall_{p_\ell \in S} (v \neq \mathit{brb}[\typ].\mathsf{deliver}(\ell))$;


\medskip

\textbf{operations:}
$\mathsf{vbbBroadcast}(v)$ \label{ln:bvBradcast}\textbf{do} $\mathit{brb}[\init][i].\mathsf{broadcast}((i,v))$\;

$\mathsf{vbbDeliver}(k)$ \label{ln:vbbDeliver} \Begin{
	
	\lIf{$\mathit{brb}[\initE][k]=\bot \land \mathit{brb}[\validE][k] \neq \bot$\label{ln:conTestBRB}}{\Return{$\blitza$}} 
	
	\lIf{$\exists p_j,p_k\in\sP,\typ\in\texttt{vbbMSG} :\mathit{brb}[\typ][j].\mathsf{deliver}()=(k,\bull)\land j\neq k$\label{ln:ilegalInputtoBRB}}{\Return{$\blitza$}}
	
	\lIf{\emsA{$\mathit{brb}[\initE][k].\mathsf{deliver}()\neq(k,v) \lor \mathit{brb}[\validE][k].\mathsf{deliver}()\neq(k,x)$\label{ln:deliverKx}}}{\Return{$\bot$}}  
	
	\lElseIf{\emsA{$v\notin V \lor x \notin \{\false,\true\}$}}{\Return{$\blitza$}\label{ln:vNotBotBlitza}}
	
	\lElseIf{$x \land \mathit{vbbEq}(\validE,v)$}{\Return{$v$}\label{ln:xVBBeqRetV}}
	
	\lElseIf{$\neg x \land \mathit{vbbDiff}(\validE,v)$\label{ln:blitzaVBBdeliverIf}}{\Return{$\blitza$}\label{ln:blitzaVBBdeliver}} 
	
	\lElseIf{$\mathit{vbbEcho}(\validE)$}{\Return{$\blitza$;\label{ln:blitzaVBBecho}} 
		
		\Return{$\bot$\label{ln:vbbDeliverElad}}
}}

\smallskip

\textbf{do-forever} \Begin{
	
	\If{$\mathit{vbbEcho}(\initE)\land v \neq \bot$ \textbf{\emph{where}} $v=\mathit{brb}[\initE][i].\mathsf{deliver}()$\label{ln:brbValid}}{$\mathit{brb}[\valid][i].\mathsf{broadcast}((i, \mathit{vbbEq}(\init,v)))$}\label{ln:brbValidThen}	
	
}

\medskip

\caption{\label{alg:SSBFTVBB}Self-stabilizing Byzantine-tolerant VBB-broadcast; code for $p_i$}
\end{\algSize}
\end{algorithm*}

\Section{Self-stabilizing Byzantine-tolerant Multivalued Consensus}


\begin{algorithm*}[t!]
\begin{\algSize}

\medskip

\textbf{variables:}
$\mathit{bvO}:=\bot$; \texttt{/*} Binary-values object, $\bot$ is the post-recycling state. \texttt{*/};

$\mathit{bcO}:=\bot$; \texttt{/*} Binary consensus object, $\bot$ is the post-recycling state. \texttt{*/};

\medskip

\textbf{macros:}
$\mathit{mcEcho}() :=\exists_{S \subseteq\sP: n \mathit{-}t\leq |S|} \forall_{p_k \in S}  (\mathsf{vbbDeliver}(k)\neq \bot)$\label{ln:mcEchoExists}\;

$\sameValue()$ \label{ln:pbDef} \textbf{do return} $(\exists_{v \notin \{\bot,\blitza\}}\exists_{S' \subseteq\sP: n \mathit{-}2t\leq |S'|} \forall_{p_{k'} \in S'} (\mathsf{vbbDeliver}(k')=v)) \land (|\{ \mathsf{vbbDeliver}(k) \notin \{\bot,\blitza\}:p_k \in \sP \}|=1)$\;

\medskip

\textbf{operations:}  			
$\mathsf{propose}(v)$ \label{ln:mvcPropuseV}\textbf{do} $\mathsf{vbbBroadcast}(v)$;

$\done()$ \label{ln:mvcDone}\Begin{
	\lIf{$\mathit{bcO} = \bot \lor \mathit{bcO}.\bcdone() = \bot$\label{ln:notReady}}{\Return{$\bot$}}
	
	\lElseIf{$\neg \mathit{bcO}.\bcdone()$\label{ln:normalReturnAndConsis}}{\Return{$\blitza$}}
	
	\lElseIf{$\exists_{v \notin \{\bot,\blitza\}} \exists_{S' \subseteq\sP: n \mathit{-}2t\leq |S'|} \forall_{p_{k'} \in S'} (\mathsf{vbbDeliver}(k')=v)$\label{ln:defultReturnIf}}{\label{ln:defultReturn}\Return{$v$}} 
	
	\lElseIf{$\emsA{\mathit{mcEcho}() \lor \True \notin\mathit{bvO}.\binValues()}$\label{ln:normalReturnAndConsisT}}{\Return{$\blitza$}}
	
	\Return{$\bot$;\label{ln:mvcDoneElad}}
}		

\smallskip

\textbf{do-forever} \Begin{
	
	\If{$\emsA{\mathit{mcEcho}()}$\label{ln:mcEcho}} {
		
		\lIf{$\mathit{bcO}= \bot$\label{ln:bcOpropuse}} {$\mathit{bcO}.\mathsf{propose}(\sameValue())$}
		
		$\mathit{bvO}.\mathsf{broadcast}(\sameValue())$\label{ln:bcOpropuseTest}\;
		
	}
	
}

\medskip

\caption{\label{alg:consensus}Self-stabilizing Byzantine- and intrusion-tolerant multivalued consensus via VBB-broadcast; code for $p_i$}
\end{\algSize}
\end{algorithm*}

\emsA{Algorithms~\ref{alg:consensus} and~\ref{alg:SSBFTVBB} present our self-stabilizing Byzantine- and intrusion-tolerant solution to the problem of multivalued consensus using an SSBFT VBB-broadcast solution.} 
\ems{They are obtained from algorithms~\ref{alg:vbbBroadcast} and~\ref{alg:reductionNon} via code transformation and the addition of necessary consistency tests (sections~\ref{sec:vbbInv} and~\ref{sec:mvcInv}).}
Note that the line numbers of \emsA{algorithms~\ref{alg:SSBFTVBB} and~\ref{alg:consensus}} continue the ones of Algorithm~\ref{alg:reductionNon}. 

\ems{\Subsection{SSBFT VBB-broadcast}
The operation $\mathsf{vbbBroadcast}(v)$ allows the invocation of a VBB-broadcast instance with the value $v$. Node $p_i$ VBB-delivers messages from $p_k$ via $\mathsf{vbbDeliver}_i(k)$.} 

\Subsubsection{Types, constants, and variables}
\ems{We define the \emsA{phase types of} $\texttt{vbbMSG} :=\{\init, \valid\}$ (line~\ref{ln:types}) and the array $\mathit{brb}[\texttt{vbbMSG}][\sP]$ for holding BRB objects, which disseminate VBB-broadcast messages, \ie $\mathit{brb}[\init]$ and $\mathit{brb}[\valid]$ store the information that VBB-broadcast disseminate of $\mathrm{INIT}()$, and \respectivelyC $\mathrm{VALID}()$ messages in Algorithm~\ref{alg:vbbBroadcast}. 
\emsA{After the recycling of these objects (Section~\ref{sec:arch}) or before they ever become active, they each have the value $[\bot,\ldots,\bot]$.} 
They become active via the invocation, say by $p_i$, of $\mathit{brb}_i[\bull][i].\mathsf{broadcast}(v)$ (which also leads to $\mathit{brb}_i[\bull][i]\neq\bot$) or the arrival of BRB protocol messages, say, from $p_j$ (which leads to $\mathit{brb}_i[\bull][j]\neq\bot$).} \emsA{We clarify that once a BRB message arrives, a call to $\mathit{brb}_i[\bull][j].\mathsf{delivery}()$ can retrieve the arriving message.}  

\ems{\Subsubsection{The $\mathsf{vbbBroadcast}()$ operation (lines~\ref{ln:bvBradcast} and~\ref{ln:brbValid})}
As in line~\ref{ln:brbBradcast0} in Algorithm~\ref{alg:vbbBroadcast}, the invocation of $\mathsf{vbbBroadcast}(v)$ (line~\ref{ln:bvBradcast}) leads to the invocation of $\mathit{brb}[\init][\bull].\mathsf{broadcast}(v)$.}
\ems{Algorithm~\ref{alg:consensus} uses line~\ref{ln:brbValid} for implementing the logic of  lines~\ref{ln:brbBradcast0wait} and~\ref{ln:brbBradcast1} in Algorithm~\ref{alg:vbbBroadcast} \emsA{as well as the consistency test of item~\ref{itm:nextPhase} in Section~\ref{sec:vbbInv}; that case of moving from phase $\init$ to $\valid$.} In detail, the macro $\mathit{vbbEcho}(\emsA{\typ})$ returns $\true$ whenever the BRB object $\mathit{brb}[\typ]$ has a message to BRB-deliver from at least $n-t$ different nodes. Thus, $p_i$ can ``wait'' for BRB deliveries from at least $n-t$ distinct nodes by testing $\mathit{vbbEcho}_i(\init)\land v \neq \bot$, where $v=\mathit{brb}_i[\init][i].\mathsf{deliver}()$. Also, the macro $\mathit{vbbEq}()$ is a detailed implementation of the function $\mathit{equal}()$ used by Algorithm~\ref{alg:vbbBroadcast}.}



\ems{\Subsubsection{The $\mathsf{vbbDeliver}()$ operation (lines~\ref{ln:vbbDeliver} and~\ref{ln:vbbDeliverElad})}
The proposed $\mathsf{vbbDeliver}()$ (lines~\ref{ln:vbbDeliver} to~\ref{ln:vbbDeliverElad}) is based on lines~\ref{ln:vbbBackground} and~\ref{ln:vbbDeliverA} in Algorithm~\ref{alg:vbbBroadcast} together with a number of consistency tests, \emsA{which are listed in Section~\ref{sec:vbbInv}.}} 

\ems{The first if-statement (line~\ref{ln:conTestBRB}) considers the (inconsistent) case in which the state of node $p_i$ encodes the fact that VBB-broadcast of the $\mathrm{VALID}()$ message occurred before the one of $\mathrm{INIT}()$ message.}
\emsA{This matches item~\ref{itm:ValidWithoutInit} in Section~\ref{sec:vbbInv}.}

The second, third, and fourth if-statements (lines~\ref{ln:conTestBRB} to~\ref{ln:blitzaVBBdeliver}) implement the logic of lines~\ref{ln:vbbWaitValidINIT} to~\ref{ln:ifXtrueElse} in Algorithm~\ref{alg:vbbBroadcast}.
Similar to line~\ref{ln:vbbWaitValidINIT} in Algorithm~\ref{alg:vbbBroadcast}, $x_i$ is the value that line~\ref{ln:deliverKx} uses for holding the value of that $p_i$ BRB-delivers from $p_k$ via the BRB object $\mathit{brb}_i[\valid]$. 
\ems{Also, the macro $\mathit{vbbDiff}()$ is a detailed implementation of the function $\mathit{differ}()$ used by Algorithm~\ref{alg:vbbBroadcast}.}
\emsA{We clarify that lines~\ref{ln:ilegalInputtoBRB} and~\ref{ln:vNotBotBlitza} return $\blitza$ when the delivered BRB message is ill-formatted. By that, they fit the consistency test of item~\ref{itm:badFormat} in Section~\ref{sec:vbbInv}; the case of transitioning from phase $\valid$ to operation completion.} 

\ems{The fifth if-statement (line~\ref{ln:blitzaVBBecho}) considers the case in which the variable $x_i$ is corrupted. Thus, there is a need to return the error symbol, $\blitza$. This happens when $p_i$ VBB-delivered $\mathrm{VALID}()$ messages from at least $n\mathit{-}t$ different nodes, but none of the if-stamemnt conditions in lines~\ref{ln:conTestBRB} to~\ref{ln:blitzaVBBdeliverIf}  hold.} \emsA{This fits the consistency test of item~\ref{itm:nextPhase} in Section~\ref{sec:vbbInv}, which requires eventual completion even in the presence of transient faults.}

\Subsection{SSBFT multivalued consensus}
\ems{The invocation of the $\mathsf{propose}(v)$ operation VBB-broadcasts $v$. Node $p_i$ VBB-delivers messages from $p_k$ via the $\done_i()$ operation. 
The logic of lines~\ref{ln:mvcPropuse} and~\ref{ln:mvcElse} in Algorithm~\ref{alg:reductionNon} is implemented by lines~\ref{ln:mvcPropuseV} to~\ref{ln:bcOpropuse} in Algorithm~\ref{alg:consensus}.}


\emsA{Algorithm~\ref{alg:consensus}'s state includes the SSBFT BV object, $\mathit{bvO}$, and SSBFT Binary object, $\mathit{bcO}$. 
Each has the post-recycling value of $\bot$, \ie when $\mathit{bvO}=\bot$ (or $\mathit{bcO}=\bot$) the object is said to be inactive.
They become active upon invocation and complete according to their specifications (sections~\ref{sec:BRBext} and~\ref{sec:sefBVbrodcast}, \respectively).}

\ems{Just like in lines~\ref{ln:mvcPropuse} and~\ref{ln:mvcESTsend} in Algorithm~\ref{alg:reductionNon}, the invocation of $\mathsf{propose}(v)$ (line~\ref{ln:mvcPropuseV}) leads to the VBB-broadcast of $v$.}

\ems{The logic of lines~\ref{ln:mvcWait} and~\ref{ln:mvcIf} in Algorithm~\ref{alg:reductionNon} is implemented by line~\ref{ln:bcOpropuse}.
In detail, if $\mathit{bcO}$ is in its post-recycling state (Section~\ref{sec:arch}) and there are ready-to-be-delivered VBB messages from at least $n-t$ different nodes, Algorithm~\ref{alg:consensus} proposes the returned value from $\sameValue()$. 
Note that the macro $\sameValue()$ (line~\ref{ln:mvcLet}) implements that predicate $\sameValue()$ (line~\ref{ln:pbDef} in Algorithm~\ref{alg:reductionNon}).}
\emsA{Line~\ref{ln:bcOpropuseTest} facilitates the implementation of the consistency test (Section~\ref{sec:mvcInv}) by BV-broadcasting the returned value $\sameValue()$. 
This way it is possible to detect the case in which all correct nodes BV-broadcast a value that is, due to a transient fault, different than $\mathit{bcO}$'s decided one. We explain how this can be done when we discuss line~\ref{ln:normalReturnAndConsisT}.}

\ems{The operation $\done()$ (lines~\ref{ln:mvcDone} to~\ref{ln:mvcDoneElad}) returns the decided value, which lines~\ref{ln:mvcThen} and~\ref{ln:mvcElse} implement in Algorithm~\ref{alg:reductionNon}. 
Since $\done()$ is a query-based operation (Section~\ref{sec:BRBext}), line~\ref{ln:notReady} considers the case in which the decision has yet to occur, \ie it returns the $\bot$-value. 
Line~\ref{ln:normalReturnAndConsisT} considers the case that line~\ref{ln:mvcIf} in Algorithm~\ref{alg:reductionNon} deals with and returns the error symbol, $\blitza$. 
Line~\ref{ln:defultReturn} implements line~\ref{ln:mvcElse} in Algorithm~\ref{alg:reductionNon}. 
\emsA{Line~\ref{ln:normalReturnAndConsisT} performs a consistency test for the case in which there are VBB-deliveries from at least $n-t$ different nodes and yet the predicate $\sameValue()$ of all correct nodes does not hold, according to the values delivered via the BV-broadcast.
This deals with the consistency test described in Section~\ref{sec:mvcInv}.}
Line~\ref{ln:mvcDoneElad} deals with the case in which none of the conditions of the if-statements above (lines~\ref{ln:notReady} to~\ref{ln:defultReturn}) hold, and thus, $\bot$ needs to be returned.}



\Section{Correctness}
We provide correctness proof for algorithms~\ref{alg:SSBFTVBB} and~\ref{alg:consensus}. The proof is organized as follows. 
%
%
%
%
%
For every layer, \ie VBB-broadcast and multivalued consensus, we provide proof of completion (theorems~\ref{thm:vbbTerminate} and \respectivelyC~\ref{thm:mvcTerminate}) before demonstrating the closure properties (theorems~\ref{thm:vbbClousre}, and \respectivelyC~\ref{thm:mvcClousre}), which \ems{show the satisfaction of the requirements of every layer.} The main difference between the completion and the closure proofs is that the latter considers post-recycling starting system states (Section~\ref{sec:arch}) and  complete (\ie proper) invocation of operations. \emsA{Due to the page limit, some of the proof details appear in the Appendix.}

\remove{

\begin{definition}[Active nodes and complete invocation of operations] 
\label{def:consistent}
%
%
We use the term \emph{active} for node $p_i \in \sP$ when referring to the case of $\exists \typ \in \texttt{vbbMSG} : \textit{brb}_i[\typ]\neq \bot$.
Suppose that during an execution $R$ that starts in a system state that is post-recycling, every correct node $p_i$ invokes \ems{VBB-broadcast (and multivalued consensus) exactly once. 
	In this case, we say that $R$ includes a \emph{complete invocation} of VBB-broadcast (and \respectivelyC multivalued consensus).}
%
\end{definition}

Note the term active (Definition~\ref{def:consistent}) does not distinguish between nodes that are active \ems{due to transient faults and legitimate invocations of $\mathsf{vbbBroadcast}()$ or} $\mathsf{propose}()$.

} 




\remove{
\begin{lemma}[BRB-no-duplicity]
\label{thm:brbDuplicity}
BRB-no-duplicity holds.
\end{lemma}
\renewcommand{\lemcnt}{\ref{thm:brbDuplicity}}
\begin{lemmaProof}
Suppose that two correct nodes, $p_i$ and $p_j$, BRB-deliver $\langle j,m \rangle$ and $\langle j,m' \rangle$, \respectivelyP The proof needs to show that $m = m'$.
If $p_i$ BRB-delivers $\langle j,m \rangle$, it received $\mathrm{MSG}(\mathit{mJ}):\mathit{mJ}[\typ][\ready]\supseteq \{(j,m)\}$ from $(2t\mathit{+}1)$ different processes, and hence from at least one non-faulty process. Similarly, $p_j$ received $\mathrm{MSG}(\mathit{mJ}):\mathit{mJ}[\typ][\ready]\supseteq \{(j,m')\}$ from at least one non-faulty process. It follows from Claim~\ref{thm:brbDuplicity} that all correct nodes broadcast $\mathrm{MSG}(\mathit{mJ}):\mathit{mJ}[\typ][\ready]\supseteq \{(j,v)\}$, from which we conclude that $m = v$ and $m' = v$.
\end{lemmaProof}
} 

\Subsection{Completion of VBB-broadcast}
\label{sec:CompletionVBB}
\emsA{As explained in Section~\ref{sec:arch}, the availability of the object recycling mechanism allows us to focus on the completion property when demonstrating recovery after the occurrence of the last transient fault. Once all (possibly corrupted) objects have completed their tasks, the mechanism will bring these objects to their post-recycling state from which the closure property can be demonstrated (Section~\ref{sec:VBBclosure}).} 

\begin{theorem}[VBB-completion]
\label{thm:vbbTerminate}
Let $R$ be an Algorithm~\ref{alg:consensus}'s execution in which all correct nodes eventually invoke $\mathsf{vbbBroadcast}()$. 
Eventually, $\forall_{i,j \in \Correct} \mathsf{vbbDeliver}_j(i) \neq \bot$. 
\end{theorem}	
\renewcommand{\thmcnt}{\ref{thm:vbbTerminate}}
\begin{theoremProof}
\ems{Let $ i \in \Correct$.} 
Suppose either $p_i$ VBB-broadcasts $m$ in $R$ or $\exists \typ \in \texttt{vbbMSG}: \mathit{brb}_j[\typ][i] \neq \bot$ holds in $R$'s starting state. 
We demonstrate that all correct nodes VBB-deliver $m' \neq \bot$ from $p_i$ \ems{by considering all the if-statements in lines~\ref{ln:conTestBRB} to~\ref{ln:blitzaVBBecho} and showing that eventually one of the if-statements in lines~\ref{ln:conTestBRB} and~\ref{ln:vNotBotBlitza} to~\ref{ln:blitzaVBBecho} holds.}


\medskip
\noindent \ems{\noindent \textbf{Argument 1.} \emph{Suppose $\exists \typ \in \texttt{vbbMSG}: \mathit{brb}_j[\typ][i] \neq \bot$. $\forall_{k,\ell \in \Correct} : \mathit{brb}_k[\typ][\ell] \neq \bot$ holds eventually.~~}
The argument is implied directly from BRB-completion-1 and BRB-completion-2.}

\medskip
\noindent \ems{\textbf{Argument 2.} \emph{Suppose that throughout $R$, the if-statement condition in line~\ref{ln:conTestBRB} does not hold. Eventually,  $\mathit{brb}_i[\validE][i] \neq \bot$ holds.~~}
By the assumption that all correct nodes are active eventually, the BRB properties (Definition~\ref{def:prbDef}) and that there are at least $(n \mathit{-} t)$ correct nodes, the if-statement condition in line~\ref{ln:brbValid} holds eventually. 
Then, $p_i$ makes sure that, eventually, the second clause in the condition of the if-statement in line~\ref{ln:brbValid} does not hold by invoking $\mathit{brb}_i[\valid][i]\mathsf{broadcast}(\bull)$ and BRB-completion.}

\medskip
\noindent \ems{\textbf{Argument 3.} \emph{Suppose $\mathit{brb}_i[\validE][i] \neq \bot$ holds in $R$'s starting state. Eventually, either the if-statement condition in line~\ref{ln:conTestBRB} holds or the one in line~\ref{ln:deliverKx} cannot hold.~~}
The proof is directly implied by the code of Algorithm~\ref{alg:consensus}.}

\medskip
\noindent \textbf{Argument 4.} \emph{Eventually, $\mathsf{vbbDeliver}_j(i) \neq \bot$ holds.~~}
\ems{Suppose that none of the if-statements in lines~\ref{ln:conTestBRB} to~\ref{ln:ilegalInputtoBRB} and~\ref{ln:vNotBotBlitza} to~\ref{ln:blitzaVBBdeliver} ever hold. Due to $\mathit{vbbEcho}()$'s definition (line~\ref{ln:auxVSsP2tpkS}), the BRB properties (Definition~\ref{def:prbDef}), the presence of at least $n\mathit{-}t$ correct and eventually active nodes, and arguments (1) to (3), the if-statement condition in line~\ref{ln:blitzaVBBecho} eventually holds.}
\end{theoremProof}

\Subsection{Closure of VBB-broadcast}
\label{sec:VBBclosure}
\ems{Theorem~\ref{thm:vbbClousre}'s proof mostly follows the arguments used for showing MR's correctness. But, there is a need to show that none of the consistency tests causes false error indications.}


\begin{theorem}[VBB-Closure]  
\label{thm:vbbClousre}
Let $R$ be \ems{an} Algorithm~\ref{alg:consensus}'s execution in which all correct nodes eventually invoke $\mathsf{vbbBroadcast}()$ \ems{and $R$'s starting system state is post-recycling (Section~\ref{sec:arch}). 
Execution $R$ satisfies the VBB requirements (Section~\ref{sec:specVVB}).}
\end{theorem}

\renewcommand{\thmcnt}{\ref{thm:vbbClousre}}
\begin{theoremProof}
VBB-completion holds (Theorem~\ref{thm:vbbTerminate}).

\begin{lemma}[VBB-uniformity]
\label{thm:vbbUniformity}
%
VBB-uniformity holds. 
\end{lemma}	
\renewcommand{\lemcnt}{\ref{thm:vbbUniformity}}
\begin{lemmaProof}
\ems{
	Let $ i \in \Correct$. 
	Suppose $p_i$ VBB-delivers $m' \in \{m,\blitza\}$ from a (possibly faulty) $p_j \in \sP$. 
	The proof shows that all the correct nodes VBB-deliver the same message $m'$ from $p_j$.
	Since $R$ is post-recycling and $p_i$ VBB-delivers $m'$ from node $p_j$, the condition $\mathit{brb}_i[\init][j]=\bot \land \mathit{brb}_i[\valid][j] \neq \bot$ (of the if-statement in line~\ref{ln:conTestBRB}) cannot hold and eventually $(\mathit{brb}_i[\init][j].\mathsf{deliver}()=(j,v_{j,i}) \land \mathit{brb}_i[\valid][j].\mathsf{deliver}()$ $=(j,x_{j,i}))$ (line~\ref{ln:deliverKx}) must hold due to BRB-completion-1 and since all correct nodes are active eventually. 
	Also, $\mathit{brb}_k[\init][j]=\bot \land \mathit{brb}_k[\valid][j] \neq \bot$ cannot hold. And,
	$(\mathit{brb}_k[\init][j].\mathsf{deliver}()=(j,v_{j,k}) \land \mathit{brb}_k[\valid][j].\mathsf{deliver}()$ $=(j,x_{j,k}))$ holds eventually, such that $v_{j,i}=v_{j,k}$ and $x_{j,i}=x_{j,k}$.
	This is because $R$ starts in a post-recycling system state, BRB-no-duplicity, and BRB-completion-2, which means that every correct node $p_k$ eventually BRB-delivers the same messages that $p_i$ delivers. 
	Due to similar reasons, depending on the value of $x_{j,i}=x_{j,k}$, the condition of the if-statement in lines~\ref{ln:xVBBeqRetV} or~\ref{ln:blitzaVBBdeliver} must hold. \Ie $p_k$ eventually VBB-delivers the same value as $p_i$ does.}
\end{lemmaProof}

\begin{lemma}[VBB-obligation]
\label{thm:vbbObligation}
VBB-obligation holds.
\end{lemma} 
\renewcommand{\lemcnt}{\ref{thm:vbbObligation}}
\begin{lemmaProof}
Suppose all correct nodes, $p_j$, VBB-broadcast the same value $v$. The proof shows that every correct node, $p_i$, VBB-delivers $v$ from $p_j$. Since every correct node eventually invokes $\mathsf{vbbBroadcast}(v)$, node $p_j$ invokes $\mathit{brb}_j[\init][j].\mathsf{broadcast}((j,v))$ (line~\ref{ln:bvBradcast}). Thus, the if-statement condition in line~\ref{ln:brbValid} holds eventually for any correct node $p_i$, \ie it is true \ems{since $\exists_{S \subseteq\sP: n \mathit{-}t\leq |S|}: \forall_{p_k \in \sP}: \mathit{brb}_i[\init][k].\mathsf{deliver}()\neq \bot$ holds eventually due BRB-completion-1.} Also, there are at least $(n\mathit{-}2t)$ appearances of $(\bull,v)$ in the multi-set $\{\mathit{brb}_i[\init][k].\mathsf{deliver}()\}_{p_k \in \sP}$. Thus, $p_i$ BRB-broadcasts the message $(\valid,(i,\true))$ (line~\ref{ln:brbValidThen}). And, for any $k,\ell \in \Correct$, \ems{$\mathit{brb}_k[\valid][\ell].\mathsf{deliver}()=(j,\true)$ holds} eventually (due to BRB-validity and BRB-completion-1). This means that\ems{, eventually, none of the if-statement conditions at lines~\ref{ln:conTestBRB} to~\ref{ln:vNotBotBlitza} holds.} However, the if-statement condition \ems{in line~\ref{ln:xVBBeqRetV}} must hold eventually and only for the value $v$. \ems{Then,} every correct node, $p_k$, VBB-delivers $v$ as the value VBB-broadcast by $p_j$.
\end{lemmaProof}

\begin{lemma}[VBB-justification]
\label{thm:vbbJustification}
VBB-justification holds.
\end{lemma} 

\renewcommand{\lemcnt}{\ref{thm:vbbJustification}}
\begin{lemmaProof}
Let $i \in \Correct$. Suppose $p_i$ VBB-delivers $m \notin \{\bot,\blitza\}$ in step $a_i \in R$. 
The proof shows that a correct node, $p_j$, invokes $\mathsf{vbbBroadcast}_j(v):m=(j,v)$ in $a_j \in R$, such that $a_j$ appears in $R$ before $a_i$. 
\ems{Since $m \notin \{\bot,\blitza\}$, the predicates $(\mathit{brb}_i[\init][j].\mathsf{deliver}()=(j,v) \land \mathit{brb}_i[\valid][k].\mathsf{deliver}()=(k,x))$ (line~\ref{ln:deliverKx}) and $x \land \mathit{vbbEq}(\valid,v)$ (line~\ref{ln:xVBBeqRetV}) hold, because line~\ref{ln:xVBBeqRetV} is the only line in $\mathsf{vbbDeliver}()$'s code that returns a value that is neither $\bot$ nor $\blitza$ and it can only do so when the if-statement condition in line~\ref{ln:deliverKx} does not hold. 
	%
	%
	Since $\mathit{vbbEq}_i(\valid,v)$ holds and $n \mathit{-} 2t \geq t \mathit{+} 1$, at least one correct node, say, $p_j$ that had BRB-broadcast $v$ (both for the $\init$ and $\valid$ phases in lines~\ref{ln:bvBradcast}, and \respectivelyC~\ref{ln:brbValidThen}), because $R$ starts in a post-recycling state and by Theorem~\ref{thm:vbbTerminate}'s Argument (2). Thus, $a_j$ appears in $R$ before $a_i$.}
\end{lemmaProof}
\end{theoremProof}

\Subsection{Completion of multivalued consensus}
\emsA{As explained  (sections~\ref{sec:arch} and~\ref{sec:CompletionVBB}), we demonstrate recovery from transient faults by demonstrating completion (due to the availability of the recycling mechanism).}

\begin{theorem}[BC-completion]
\label{thm:mvcTerminate}
Let $R$ be an Algorithm~\ref{alg:consensus}'s execution in which all correct nodes eventually invoke $\mathsf{propose}()$. The BC-completion property holds during $R$.
\end{theorem}	
\renewcommand{\thmcnt}{\ref{thm:mvcTerminate}}
\begin{theoremProof}
The proof shows that every correct node decides eventually, \ie $\forall i \in \Correct: \done_i() \neq \bot$.

\begin{lemma}
\label{thm:mainMVCTermB}
Eventually, $\done_i()$ cannot return $\bot$ due to the if-statement in line~\ref{ln:notReady}.
\end{lemma}
\renewcommand{\lemcnt}{\ref{thm:mainMVCTermB}}
\begin{lemmaProof}
Any correct node, $p_i$, makes sure that $\mathit{bcO}_i\neq \bot$, say, by invoking $\mathit{bcO}_i.\mathsf{propose}()$ (line~\ref{ln:bcOpropuse}).
This is due to the assumption that all correct nodes are eventually active, the definition of $\mathsf{propose}()$ (line~\ref{ln:mvcPropuseV}), VBB-completion, and the \ems{presence of} at least $(n\mathit{-}t)$ correct nodes, which implies that $\exists_{S \subseteq\sP: n \mathit{-}t\leq |S|} \forall_{p_k \in S} \, \mathsf{vbbDeliver}(k)\neq \bot$ holds eventually and the if-statement condition in line~\ref{ln:bcOpropuse} holds whenever $\mathit{bcO}_i= \bot$. 
Eventually $\mathit{bcO}_i.\bcdone() \neq \bot$ (by the completion property of Binary consensus). 
Thus, $\done_i()$ cannot return $\bot$ due to the if-statement in line~\ref{ln:notReady}.
\end{lemmaProof}

\medskip

If $\done_i()$ returns due to the if-statement in lines~\ref{ln:normalReturnAndConsis} to~\ref{ln:normalReturnAndConsisT}, then $\done_i()\neq \bot$ is straightforward.
Therefore, the rest of the proof focuses on showing that eventually one of these three if-statement conditions must hold and thus the last return statement (of $\bot$ in line~\ref{ln:mvcDoneElad}) cannot occur, see Lemma~\ref{thm:mainMVCTerm}.

\begin{lemma}
\label{thm:mainMVCTerm}
Suppose that, for any correct node, $p_i$, the if-statement conditions in lines~\ref{ln:normalReturnAndConsis} and~\ref{ln:normalReturnAndConsisT} never hold in $R$.	
Eventually, the if-statement conditions in line~\ref{ln:defultReturnIf} holds.
\end{lemma}
\renewcommand{\lemcnt}{\ref{thm:mainMVCTerm}}
\begin{lemmaProof}
By VBB-completion, $\mathit{mcEcho}_i()$ (line~\ref{ln:mcEchoExists}) must hold eventually since there are $n-t$ correct and eventually active nodes. 
Thus, by the lemma assumption that the if-statement conditions in line~\ref{ln:normalReturnAndConsisT} never hold in $R$, we know that, for any correct node $p_i$, eventually $\True \in\mathit{bvO}_i.\binValues()$ holds, due to the proporties of BV-broadcast (Section~\ref{sec:sefBVbrodcast}).
Thus, there is at least one correct node, $p_j$, for which $\sameValue_j()=\True$ when BV-broadcasting in line~\ref{ln:bcOpropuseTest}.
By VBB-uniformity, the if-statement condition in line~\ref{ln:defultReturnIf} must hold eventually for every correct node $p_i$.
\end{lemmaProof}
\end{theoremProof}

\Subsection{Closure of multivalued consensus}

Theorem~\ref{thm:mvcClousre}'s proof mostly follows the arguments used for showing MR's correctness. But, there is a need to show that none of the consistency tests causes false error indications.


\begin{theorem}[MVC closure]
\label{thm:mvcClousre}
Let $R$ be an Algorithm~\ref{alg:consensus}'s execution that starts in a post-recycling system state and in which all correct nodes eventually invoke $\mathsf{propose}()$.
The MVC requirements hold during $R$.
\end{theorem}

\renewcommand{\thmcnt}{\ref{thm:mvcClousre}}
\begin{theoremProof}
	BC-completion holds due to Theorem~\ref{thm:mvcTerminate}. 
	\begin{lemma}
		\label{thm:BCagreement}
		The BC-agreement property holds.
	\end{lemma}	
	\renewcommand{\lemcnt}{\ref{thm:BCagreement}}
	\begin{lemmaProof}
		\emsA{We show} that no two correct nodes decide differently. 
		For every correct node, $p_i$, $\mathit{bcO}_i.\bcdone() \neq \bot$ holds eventually (Theorem~\ref{thm:mvcTerminate}). By the agreement and integrity properties of Binary consensus, $\mathit{bcO}_i.\bcdone() = \false$ implies BC-agreement (line~\ref{ln:normalReturnAndConsis}).  
		
		Suppose $\mathit{bcO}_i.\bcdone() = \true$. \emsA{The proof is implied since there is no correct node, $p_i$, and  (faulty or correct) node $p_k$} for which there is a value $w \notin \{\bot,\blitza,v\}$, such that $\mathsf{vbbDeliver}_i(k)=w$. This is due to $n \mathit{-} 2t \geq t + 1$ and $\sameValue()$'s second clause (line~\ref{ln:pbDef}), which requires $v$ to be unique.
	\end{lemmaProof}
	
	\begin{lemma}
		\label{thm:MVC-validity}
		The BC-validity property holds.
	\end{lemma}	
	\renewcommand{\lemcnt}{\ref{thm:MVC-validity}}
	\begin{lemmaProof}
		Suppose that all the correct nodes propose the same value, $v$. The proof shows that $v$ is decided. Since all correct nodes propose $v$, we know that $v$ is validated (VBB-obligation). Also, all correct nodes VBB-deliver $v$ from at least $n \mathit{-} 2t$ different nodes (VBB-completion). Since $n \mathit{-} 2t > t$, value $v$ is unique.
		This is because no value $v'$ can be VBB-broadcast only by faulty nodes and still be validated (VBB-justification). Thus, the non-$\bot$ values that correct nodes can VBB-deliver 
		%
		%
		%
		are $v$ and $\blitza$. This means that $\forall i \in \Correct: \sameValue_i()=\true$, $\mathit{bcO}_i.\bcdone()=\true$ (Binary consensus validity), and correct nodes decide $v$.
	\end{lemmaProof}
	
	
	\begin{lemma}
		\label{thm:MVC-no-intrusion}
		The BC-no-intrusion property holds.
	\end{lemma}	
	\renewcommand{\lemcnt}{\ref{thm:MVC-no-intrusion}}
	\begin{lemmaProof}
		Suppose $w\neq\blitza$ is proposed only by faulty nodes. The proof shows that no correct node decides $w$. By VBB-justification, no $p_i : i \in \Correct$ VBB-delivers $w$. 
		
		Suppose that $\mathit{bcO}_i.\bcdone() \neq \true$. Thus, $w$ is not decided due the if-statement \ems{line~\ref{ln:normalReturnAndConsis}.} Suppose that $\mathit{bcO}_i.\bcdone()=\true$. There must be a node $p_j$ for which $\sameValue_j()=\true$. 
		%
		%
		%
		\Ie $v$ is decided due to the if-statement \ems{in line~\ref{ln:defultReturnIf}} and since there are at least $n \mathit{-} 2t$ VBB-deliveries of $v$. 
		\ems{Note that the if-statement condition in line~\ref{ln:normalReturnAndConsisT} cannot hold during $R$ since $R$ starts in a post-recycling system state (as well as due to lines~\ref{ln:bcOpropuse} and~\ref{ln:bcOpropuseTest}, which use the same input value from $\sameValue()$).} 
		This implies that $w\neq v$ cannot be decided since $n \mathit{-} 2t > t$.
	\end{lemmaProof}
\end{theoremProof}

\Section{Discussion}
To the best of our knowledge, this paper presents the first self-stabilizing Byzantine- and intrusion-tolerant algorithm for solving multivalued consensus in asynchronous message-passing systems. This solution is devised by layering broadcast protocols, such as Byzantine reliable broadcast, Binary-values broadcast, and validated Byzantine broadcast. Our solution is based on a code transformation of existing (non-self-stabilizing) BFT algorithms into the proposed self-stabilizing Byzantine-tolerant algorithm. This transformation is achieved via careful analysis of the effect that arbitrary transient faults can have on the system's state as well as via rigorous proof for demonstrating consistency regaining and completion. We hope that the proposed solution and studied techniques can facilitate the design of new building blocks, such as state-machine replication, for the Cloud and distributed ledgers.




\part*{Appendix}


\begin{table}[h!]
	\begin{tabular}{|l|l|}
		\hline
		\textbf{Notation} & \textbf{Meaning} \\ \hline \hline
		BFT &    (non-self-stabilizing) Byzantine fault-tolerant     \\ \hline
		BRB &    Byzantine-tolerant Reliable Broadcast, \eg the SSBFT one in~\cite{DBLP:journals/corr/abs-2201-12880}     \\ \hline
		BV-broadcast  &  Binary-values broadcast, \eg the SSBFT one in~\cite{DBLP:journals/corr/abs-2103-14649}      \\ \hline
		MR & the studied solution by Most{\'{e}}faoui and Raynal~\cite{DBLP:conf/opodis/MostefaouiR10}     \\ \hline
		SSBFT &    self-stabilizing Byzantine fault-tolerant     \\ \hline
		VBB  &  Validated Byzantine Broadcast, \eg the BFT ones in algorithms~\ref{alg:vbbBroadcast} and~\ref{alg:SSBFTVBB}   \\ \hline
	\end{tabular}
	\caption{\label{fig:Glossary}Glossary}
\end{table}


\end{document}